# Fast, Accurate, but Sometimes Too-Compelling Support: The Impact of Imperfectly Automated Cues in an Augmented Reality Head-Mounted Display on Visual Search Performance

Amelia C. Warden, Christopher D. Wickens, Daniel Rehberg, Francisco R. Ortega, and Benjamin A. Clegg

*Abstract*— While visual search for targets within a complex scene might benefit from using augmented-reality (AR) head-mounted display (HMD) technologies helping to efficiently direct human attention, imperfectly reliable automation support could manifest in occasional errors. The current study examined the effectiveness of different HMD cues that might support visual search performance and their respective consequences following automation errors. Fifty-six participants searched a 3D environment containing 48 objects in a room, in order to locate a target object that was viewed prior to each trial. They searched either unaided or assisted by one of three HMD types of cues: an arrow pointing to the target, a plan-view minimap highlighting the target, and a constantly visible icon depicting the appearance of the target object. The cue was incorrect on 17% of the trials for one group of participants and 100% correct for the second group. Through both analysis and modeling of both search speed and accuracy, the results indicated that the arrow and minimap cues depicting location information were more effective than the icon cue depicting visual appearance, both overall, and when the cue was correct. However, there was a tradeoff on the infrequent occasions when the cue erred. The most effective AR-based cue led to a greater automation bias, in which the cue was more often blindly followed without careful examination of the raw images. The results speak to the benefits of augmented reality and the need to examine potential costs when AR-conveyed information may be incorrect because of imperfectly reliable systems.

## I. INTRODUCTION

VISUAL search is ubiquitous in the workplace and everyday life [1], [2]. Examples include a driver or pilot searching the forward view for potential collision hazards [03], the industrial worker or quality control inspector searching a product for flaws [4], [5], the radiologist searching medical imagery for a tumor or fracture [6], [7], the TSA inspector searching the x-ray image for weapon [8], [9], [10], [11], [12], the computer user searching a cluttered screen for a specific file [13] or the warfighter searching a forward scene for a potentially hostile target [14], [15], [16] [17], [18]. When unaided, visual search can be time-consuming and error-prone, with a common issue being missing true targets when they are present [19]. This high human performance cost results because complex visual search is often **serial**, effortful, and inefficient, involving the inspection of each item in the search field in turn until one is encountered that is suspected to be the target [20], [21].

Challenges in both speed and accuracy mean there is great potential value to human performance of a search **cue** that can direct the searcher's attention to that object [2], [18], [10]. An object might be cued based on an automation or AI inference that that particular object is likely to be a target. The current research addresses the intersection of human performance and computer technology in cued search in two respects: (1) we examine the technology of augmented reality cueing, and (2) we consider the imperfections of automation (e.g., machine vision) that may render the AI-based cueing imperfectly reliable.

The specific contribution of this paper is that we demonstrate the differential effectiveness of three different types of head-mounted display (HMD) cueing aids in a comparison that has not been previously reported. Head-mounted displays integrate small displays or projection technology into devices such as eyeglasses, visors, or devices that are mounted onto a helmet. One fundamental advantage of an HMD, specifically in the case where it displays augmented reality (AR) information (e.g., target cueing) is that they reduce the scanning between the display and the real-world viewed through the display. This allows the system to

This research was supported by the Office of Naval Research under grant numbers N00014-21-1-2949 and N00014-21-1-2580. Dr. Peter Squire was the scientific/technical monitor.

Amelia C. Warden is with the Department of Psychology, Colorado State University, Fort Collins, CO 80523 USA (email: acwarden@colostate.edu).

Christopher D. Wickens is with the Department of Psychology, Colorado State University, Fort Collins, CO 80523 USA (email: chris.wickens@colostate.edu) .

Daniel Rehberg is with the Department of Computer Science, Colorado State University, Fort Collins, CO 80523 USA (email: Dan.Rehberg@colostate.edu).

Francisco R. Ortega is with the Department of Computer Science, Colorado State University, Fort Collins, CO 80523 USA (email: F.Ortega@colostate.edu).

Benjamin A. Clegg is with the Department of Psychology, Colorado State University, Fort Collins, CO 80523 USA (e-mail: Benjamin.clegg@colostate.edu).

visually designate a suspected target while the searcher looks directly at the search field in the real-world beyond the display.

We provide an underlying psychological model to account for the differences in effectiveness of the 3 cueing techniques. In particular, augmented reality head-mounted display (AR-HMD) cueing based on imperfect automation has been rarely examined, and in our examination, we provide a unique comparison revealing that greater effectiveness of a more realistic world-centered cue when it is correct may be offset by greater cost on the infrequent occasions when it is wrong.

## II. BACKGROUND

### A. Perfect Cueing

In theory, unaided search through a series of objects to find a target that is similar in appearance to the non-target objects will take, on average, a search time of:

$$ST = a + \frac{nT}{2} \tag{1}$$

where $a$ is a constant associated with making the response when a target is located but, importantly, it also can involve a time-consuming target confirmation process; $n$ is the number of non-target objects in the search field; and T is the time to inspect each object and determine that it is **not** a target. The division by 2 represents the fact that, on average, the target will be found after half of the objects have been inspected. This is the *serial self-terminating search* (SSTS) model [20] which has been confirmed in multiple experiments over the past 50 years (See [2] for review). There are multiple exceptions and qualifications to this pure serial search model [21], [22] but in many applied instances it provides a reasonable approximation to predicting ST [23].

Thus, a perfect cue may be able to drastically shorten this time from $(a + \frac{nT}{2})$ to $(a + T)$, as the single cued object is inspected to confirm that it is the target. A study in our lab confirmed a reduction in search time for 3D objects located around a room from 10 to 5 seconds, associated with one type of perfect cueing [24]. Here we can calculate that 4.6 seconds represented the constant $a$ in equation (1), with T = 0.33 seconds. In hostile, or dynamic environments, saving a few seconds can be of vital importance. Consider, for example, cueing the location of a potential collision hazard while driving fast or flying [25]. Such is an example of one of the many aspects of Wolfe's [21], [22] guided search model, which can shorten search time, where the cue guides attention to a single candidate for a target.

Naturally, the benefits of the cue in reducing search time will also be based on what we label the *cue effectiveness value* (CEV), which accounts for the time required to see and interpret the cue and orient attention from the cue to the cued target [2], [26], [27], [28], [29], an issue of direct interest in the current study. The CEV can be predicted and modeled by considering the human information processing operations involved in the search. One important influence on these operations is the extent to which the cue is considered *exogenous* or *endogenous* [2], [28]. An endogenous cue presents information about the cued location symbolically but does not appear at the location of the cue itself. For example, an icon cue conveys the appearance of a target but does not directly mark the location where attention should be shifted. Conversely, an exogenous cue orients the user to the location where attention should be shifted, such as a flashing highlight (presented on an HMD) near an object in the world that automatically orients attention towards the location of a target in the search scene.

Basic visual attention research reveals that reliable exogenous (egocentric) cueing is more rapidly processed (greater CEV, smaller CEV penalty) than endogenous cueing. However, while the endogenous-exogenous distinction is widely used in basic research on visual attention, exogenous cues are only effective if the target is in the forward eye field and runs the risk of capturing attention when people are not currently searching for that specific target. In the current research, we, therefore, have employed the more application-relevant terms of *world-referenced* (egocentric) versus *display-referenced* (exocentric) cueing, because this maps directly onto two philosophies of HMD cueing, the focus of our investigation.

In HMD design, world-referenced cueing capitalizes on the properties of augmented reality, to position and re-position a cue on an HMD such that it always overlays or points directly toward the target, irrespective of how the head is oriented. That is, the cue is presented on the display in world-referenced coordinates that are rapidly updated on the display as the head rotates laterally or vertically. This is one of the cue types to be evaluated in the current study, in which an AR *arrow* cue is programmed to always point toward the location of the target.

Screen-referenced exocentric or display-referenced cues are presented at the same fixed screen coordinates independent of head rotation. They can nevertheless designate in some less direct fashion that object that automation infers to be the target, either by specifying its location or its visual appearance. In our research, a location cue highlights the inferred target as depicted on a *minimap,* positioned in screen-referenced coordinates on the HMD [30], [24]. The visual appearance cue, which we call the *icon cue*, presents a picture (icon) of the inferred target, continuously visually available, again presented at fixed screen-referenced coordinates. Both of these cue types impose CEV penalties compared to the AR world-referenced arrow cue, but the penalties are of different varieties. For the minimap cue, the penalty is one of the spatial frame-of-reference transformations necessary to determine where the cued target viewed on the minimap (typically a top-down planar display) is located on the forward ego-referenced 3D view of the search field [31], [32]. For the icon cue, the penalty is inherent in the increase in N because of the many of the objects in the search field that still need to be sequentially examined to determine which one matches the icon. However, some shortening of search time is still preserved here to the extent that salient features of the target icon (e.g., a green color or large size), can discriminate it from many, although not all, of the non-target objects in the search field. That is, some salient property of the target on any particular trial may help guide the search through a reduced number of items [21], [22].

In summary, in comparing the three cue types, their search time, and hence benefit over the control condition (equation 1), is predicted to be:

$$AR\ Arrow: (a + T)\qquad(2)$$

Here there is only a single item to be searched for and a single confirmation. Because the cue unambiguously points to the inferred target, CEV is maximum (i.e., having no penalty).

$$Minimap: (a + T) + (FORT\ CEV\ penalty)\qquad(3)$$

This is identical to the predicted search time for the arrow cue, except there is now a penalty to the CEV related to the frame of reference transformation (FORT) of mental rotation from the top-down plan-view minimap to the forward 3D perspective view of the search field.

$$Icon: a' + \frac{((n-x)T)}{2}\qquad(4)$$

The icon cue differs from the arrow cue in two ways. First, as noted above, relative to the control condition, search time may be reduced if certain salient features of some targets may reduce the search field from n by a quantity $x$, as predicted by the guided search model [21]. Second, as noted in the context of the uncued search (equation 1), in all of the equations the constant ***a*** actually contains two subcomponents. One is the time to actually execute the target identification response, which is identical for all four conditions. The other, non-trivial component is the time required to confirm that an initially identified candidate is actually the target. This will involve, as participants are requested to do, a close comparison between the candidate target seen in the real world through the HMD and the representation of the target, either visually on the HMD (the icon) or in visual memory (the minimap or arrow). This confirmation time is likely to differ between the arrow and minimap on the one hand ($a$), and the icon ($a'$) on the other hand. This difference results because, for the two location cues, the comparison must be made with a degraded remembered image, due to the passage of time, of the true target; whereas for the icon, the image remains directly visible, and so invites a careful comparison of the features of the image with that of the suspected target in the real world. This may take more time, even if it also improves accuracy. While the current experiment does not contain the necessary controls to obtain precise estimates of all these terms, it is possible to gauge rough estimates.

Importantly, in an experiment examining these three types of cues [24], results showed cueing benefits of 5, 4, and 1, seconds for the arrow, minimap, and icon cues, respectively, indicating that the CEV of location cueing (i.e., the arrow and the minimap) was far more effective than that of visual appearance (i.e., the icon). However, the icon cue can also provide a benefit to search accuracy not conferred by the other types of cues: that is, the icon cue allows the searcher the capability to confirm with certainty the identity of the target, rather than relying upon the potentially degraded memory of what the target looked like. Thus, Warden and colleagues [24] found that the accuracy CEV was just as great (statistically) for the icon cue as for the arrow or minimap cue.

### B. Imperfect Cueing and Automation

In [24], the cues were always correct. But in the operational world, a cue must be guided by an automation inference as to what should be the target. Such inference might often be guided by computer vision using machine-learning that examines each object and, like the human, compares the product of this examination with a template of what a true target looks like. As with so many other functions of automation, this inference may, on occasion, be incorrect [33], [34], [35], [36], [37], [10] due to various factors such as lighting conditions, pattern complexity in the search scene, and the reliability of the automation machine-vision system. Such imperfectly reliable automation also has profound implications for the performance of the human-automation team, depending upon the degree of dependence or reliance of the human on the automation's decision/diagnosis of what is deemed important to be attended. Indeed, decades of research on human trust in imperfect automation have examined these factors [38], [34], [35], [39]. A smaller number of studies have examined it specifically in the context of visual search and target cueing (e.g., [15], [30], [36], [37], [8], [10]). However, only one study [15] appears to have done so in the specific context of AR-HMD target cueing; (although see also [30] for a closely related study).

Three general findings have emerged from the above research on human use of imperfect automation. First, there is a general tendency for humans to depend upon automation recommendation, whether this recommendation is of a course of action, a location to attend to, or a diagnostic state, without adequately examining the raw data upon which that recommendation is based [40]. This dependency is known as the ***automation bias*** [41], [42], and has the consequence that, on instances when automation fails, which occur with a frequency of $(1.0 - reliability)$, humans are also likely to err in their automation-based judgment. In visual search tasks, such automation failures may be either failing to find a true target (automation miss) or falsely classifying a non-target as a target (automation false alarm). Somewhat ironically, these failures become more likely, the more reliable automation is [43] (but see [10]), as humans may sometimes express what is called a "perfect automation schema" which is associated with high expectations of automation performance and less forgiving attitudes of automation failures than human failures [44], [45].

Secondly, as the reliability of automation does decrease, users are somewhat sensitive to this decrease (although more so to automation false alarms than misses [46]); their trust in automation is reduced and their dependence on automation is correspondingly reduced. Consequently, the extent of the automation bias is also reduced. Unlike the previous HMD cueing type comparison [24], the present study included **imperfect** automation cueing, at an error rate of 17%, a value that may indeed be typical of many automation systems in use [47], but a level of imperfect reliability that is nevertheless helpful in aiding the human-automation team performance [48].

Third, there is some evidence that the effect of cue imperfections may interact with the type of cue employed, specifically that the more realistic "compelling" AR arrow cue, while providing greater assistance when it is correct, may lead to greater human errors caused by the automation bias on the infrequent occasions when the cue is wrong. Some basis for this prediction is provided by the research examining imperfect AR target cueing [15]. In this study, the authors employed an imperfect (85% reliable) cue of a potential hostile enemy target to soldier participants either with a head-down hand-held display to signal possible target location or with an AR-HMD ego-centric arrow cue. They observed that the latter was more effective when the cue was correct, but more problematic when the cue was wrong providing a false alarm (i.e., cueing a non-target). It was as if the more immersive *world-referenced* cue amplified an automation bias.

This tendency aligns with the idea that the more immersive cue produces more *attentional tunneling* [14], [49] at the expense of examining the raw data in the natural world beyond the HMD. In support of this causal assumption, [14], [15] also observed that the AR-HMD cueing led to a decrease in the detection of other non-cued but high priority threats in the scene. In the current study, we are predicting that all three automation cues will create some degree of automation bias, but that this will be amplified with the arrow cue, under the assumption that the more compelling immersive and easy-to-use display (mitigating frame of reference transformations) will amplify dependence on that cue.

This prediction is also consistent with a meta-analysis of the aircraft HUD [50] which revealed that the HUD provided benefits in particular when HUD information was conformal (i.e., a one-to-one mapping between a spatial overlay of a real or virtual object and the far domain) with the world beyond – a concept that is in many ways analogous to augmented reality displays. But that conformal HUD provided a unique cost to detecting events in the world beyond that were not depicted on the HUD; a form of automation failure. A military study, [51] found that the more immersive and realistic 3D display, as with the AR-HMD used by [24], improved performance on tasks depending on information in the current field of view; but [14] and [51] found that this immersive 3D display created a sort of attentional tunneling, which degraded detection of events outside of that forward view. Misfud and colleagues [30] found that errors in a directional cue presented within an AR-HMD were frequently undetected by the human participants. Again, this is as if this human performance bias was amplified by augmented reality (See also [52]). Neither of these studies nor the prior HUD meta-analysis, directly replicate the manipulations or displays used in the current study; but they do point toward certain predicted findings regarding the potential costs of high display realism [52] served by imperfect information.

### C. Cue Location and Search Strategy

In addition to the effectiveness of cue type, a second issue examined in the current study is the effect of cue location within the visual field. Of course, the fundamental premise of using an HMD for attentional cueing in the first place is to reduce the information access effort required to move attention between the far domain and the displayed cue location [53]. While a head-down display, such as a tablet, was not employed for comparison in the current study (see for example [14]), our study did include a manipulation of cue location within the HMD. We hypothesized a benefit for placing the cue directly in the center of the HMD field of view (FOV) compared to positioning it at the bottom because of the reduced information access effort at the former location, as observed in our perfect cueing study [24]. However, we also recognize that any such advantage could be partially or fully offset by two factors:

(1) The increased clutter that results when a screen-referenced cue (icon or minimap) is directly superimposed on the foveal view of the outside world where all search objects can be seen when the direction of gaze is straight ahead. This offset is expected to grow as the visual complexity of the cue increases, and hence be larger with the more visually complex minimap, a finding observed in the perfect cueing study [24].

(2) The particular *search strategy* employed. A strategy that proceeds from the top-down, will likely produce a relatively greater clutter-induced compromise of target detection in the lower search field when the cue is presented downward. Performance using a strategy that proceeds from the center-outward on the other hand will be little affected by vertical cue location. We have a particular interest in the basic search strategy employed in this paradigm in the control condition, when there is no cue present.

Thus, the current experiment examines search for 3D target objects in a room, similar to that employed by [24] where participants were either unaided or assisted by an HMD cue of either of the three types described above. For half the participants, the cueing was imperfect where on 17% of the trials an object was cued that was not the target that they had studied prior to the start of the search. Based upon our review of the literature above, we hypothesized that:

$H_1$: Any cueing would provide a CEV benefit to search time as measured by response time (RT) relative to the control condition, given the long history of human factors research on cueing effectiveness.

$H_2$: This RT benefit would be maximum for the AR arrow

cue [24], reduced for the minimap cue because of its spatial transformation requirement that reduces CEV, and least for the icon cue which relied upon visual appearance rather than location information and forces some degree of time-consuming serial search, as described above. This is based upon both the prior findings of [24] and the information processing analysis of each different cue type described above in equations (2), (3) and (4).

H₃: Cueing would provide a CEV benefit for accuracy relative to the control condition with no cue; based on the history of cueing research (See [24]).

H₄: The profile of accuracy benefits across cue type would differ from that shown by RT, because of the improved accuracy CEV for the icon cue, which allows memory-free confirmation of the cued target. This hypothesis is based on the known frailties of human working memory when a current image must be compared against a remembered image, which is the case with both the arrow and minimap cue.

H₅: Imperfect cueing automation would impair overall search performance because of the contribution of high error human rate on those 17% of the trials when the cue was wrong [15]. This is based on the large data base of research on the effects of automation imperfection on human-automation team performance.

H₆: This impairment would be proportional to the benefit offered when the cue was correct This is based on the corresponding finding of the other AR-HMD cueing study available in the literature [15], [30].

H₇: The location of the cue in the center of the HMD FOV rather than downward would generally improve cueing effectiveness because of reduced information access effort, particularly for the visually simple arrow cue. Ample experimental evidence, reviewed in [53] predicts some reduction in performance when two sources of information that need to be integrated are displayed further apart.

## III. EXPERIMENT

### A. Method

*Participants.* A total of 57 participants in an introductory psychology course at Colorado State University received course credit in exchange for completing the experiment. The experiment was approved by the Institutional Review Board. Requirements to participate were that they should not be color blind. Before starting the experiment, participants gave their informed consent and completed an online version of the Ishihara color blindness test to rule out any unknown red-green deficiencies. Previous work indicates that online colorblindness tests are valid measures for screening for general color deficiencies [54], [55]. All participants had normal or corrected-to-normal vision, and all passed the electronic color-blind test.

*Apparatus and Stimuli.* Participants completed the experiment using a HoloLens 2, an AR-HMD. The HoloLens 2 is an AR optical see-through headset that overlays virtual content onto the real world. The field of view (FOV) of the device is 43° in the lateral direction and 29° in the vertical direction. The cueing aids used during the visual search task were created in Unity, version 2019.3.1f1. We systematically constructed 48 real-world objects using multicolored Mega Blocks. Each object consisted of five blocks pointing in the $\pm$ x, $\pm$ y, and + z planes and varying in color and shape. In addition, we created 8 objects that were not in the search scene but were similar to 8 objects in the search scene (foils). These foil objects differed from their real-world counterpart in two dimensions (e.g., color and shape). These foil objects were used for the imperfect reliable condition only.

*Task.* Participants completed a 180-degree visual search task using an AR-HMD. Participants were seated in a stationary chair surrounded in front by the search field (Figure 1) within which they were to search for a designated target on each trial. In all conditions, the HMD displayed an image of the target object to be searched at the beginning of each trial. Participants had a total of 5 seconds to study the image until the search began. Once the trial began, they scanned the search field for the designated object. Once they felt that they had located the target, they fixated directly on the object. A gaze-sensitive algorithm then rendered a box around the target. To make their selection, they pressed the 'A' button on an X-Box controller. If the item they selected was the correct item (e.g., the target image at the beginning of the trial matched the object selected in the real world), the response was recorded correct. Otherwise, the response was recorded as incorrect. For the imperfectly reliable condition, participants could select the practice block if they believed the target object at the beginning was not present. This was recorded as correct.

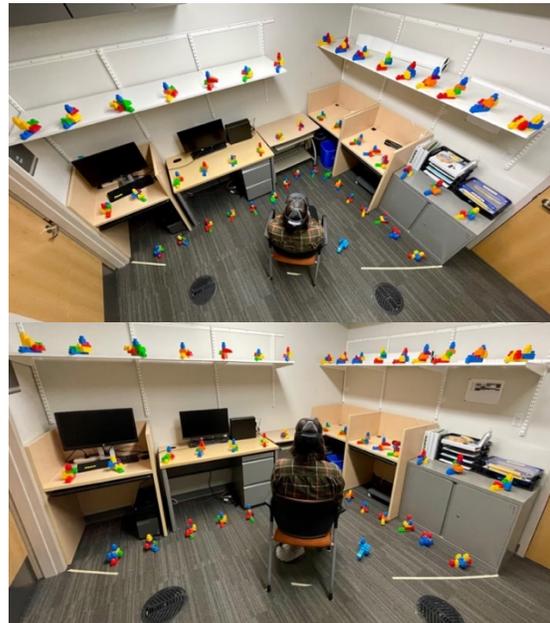

**Figure 1.** An image of the search scene depicted here from a top-down perspective (top) and a forward-facing perspective (bottom). Objects were placed three heights (shelf, table, floor) within a 180-degree search field.

Participants completed four cueing conditions during the experiment. As shown in Figure 2, participants were given three types of cueing aids: an arrow cue, minimap cue, icon cue. In addition, they completed a no cue (control) condition. The arrow cue was a green 2D AR world-reference arrow that pointed in the direction of the target as the participant turned their head. The arrow cue turned into a green circle once the participants' gaze was on the designated target (see left panel of Figure 2). The minimap cue used screen-referenced coordinates that continuously displayed the position of all objects in the search scene from a top-down bird's eye perspective. The target object's location was indicated with a yellow box around the object, as can be seen in the center panel of Figure 2. The icon cue continuously displayed the visual appearance of the target object in screen-referenced coordinates on the HMD, which can be seen in the black box to the lower left of the right panel in Figure 2. Note that the general similarity of the cued target icon with various search objects (3 of which are seen on the ledge above) requires careful comparison to judge which search object matches the icon.

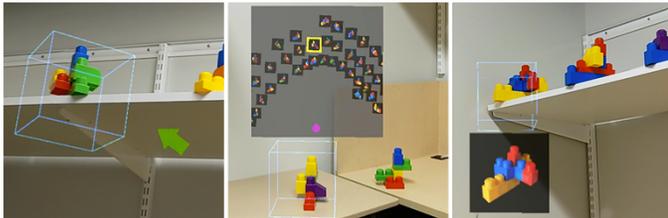

**Figure 2.** Example images of each cue with the hit boxes surrounding the object. Hit boxes denote when the participants gaze was on that specific object.

Cues were positioned in either the center or 12.6 degrees downward from the center of the AR-HMD display: this display location represents the cue-location variable in the experiment. As shown in Figure 1, the objects in the real world were distributed across the 180-degree search field, and located at three different height locations (floor, table, and shelf) that were separated by 28 inches in the vertical direction. A total of 48 objects were uniformly distributed across two halves of the room. The target object could appear at any location within this 3D array.

*Design.* The experiment was a 2 (reliability) x 3 (cue condition) x 2 (cue location) x 3 (object vertical location) mixed-subjects design, with the reliability as the between-subjects variable and cue type, cue location, and object location as the within-subjects variables. A 4th condition, no-cue was included, but this did not include cue location as a variable.

Participants were randomly assigned to one of the two reliability groups. One group of participants (N = 28) completed the 100% reliability condition for which the image of the target object always matched the object in the real world that the cueing aid located. In order to encourage participants not to rely totally on the cue, the instructions indicated that: "*The cue is always correct. However, before selecting the object that is identified by the cue, you should try to assure that the selected object is, in fact, the same as the one that you saw at the beginning of the trial. If you do not see the object in the search field, you can select the blue practice object block to indicate "object not present*.""

The second group of participants (N = 28) completed the 83% reliability condition for which the image of the target object occasionally did not match the real-world object that the cueing aid designated (i.e., the cueing aid located the incorrect object 17% of the time). On the incorrect trials, the cue instead designated the foil; that is, the object that was matched to look similar, but not identical, to the object shown at the beginning of the search. Such similarity captured the kind of error that a machine learning system might make. Participants in this condition were explicitly instructed that: "*The cueing aids may NOT BE PERFECTLY RELIABLE, meaning you may not fully rely on them. If you do not see the object in the search field, you can select the blue practice object block to indicate 'object not present'*". For the arrow and minimap cues, the spatial properties of the cue designated the object (foil) that looked like the originally viewed target. For the icon cue, the icon representation of the target object remained on the HMD, but, as with the two spatial cueing conditions, that object was not present in the search field, being replaced instead by the similarly looking object.

Participants completed 12 trials for each cue condition in a blocked, counterbalanced order; six trials when the cue was located in the center of the display and six trials when the cue was located 12.6 degrees downward from the center. These two cueing locations were blocked, and the cue conditions were counterbalanced. Participants completed 12 trials for the no cue (control) condition. The entire experiment consisted of 48 trials. Four different sets of trials were randomly generated such that each cue type was presented at one of three heights on the left and right sides of the room. For example, on the left side of the room, three trials of the 2D arrow in the center location of the display would cue the participant to three objects located at the shelf, table, and floor heights, each at one of three regions in the lateral direction of the left side of the room (see Figure 1). Of the 8 incorrect target images (2 for each cue condition) each was paired with a real-world object that was similar but not identical. These were presented at the three vertical levels on the two sides of the room based on locations within the four different sets of trials described above.

*Procedure.* Participants completed seven practice trials to familiarize themselves with each cue condition. They completed two trials for each of the three cueing aids (arrow, minimap, and icon): one with the cue in the center of the display and one with the cue 12.6 degrees downward from the center. They completed only one practice trial for the no-cue condition. They searched for the same practice block consisting of only blue Mega Blocks for all practice trials. Next, they completed the test trials where they searched for 48 different target objects for each cueing condition. After completing the visual search task, they took a Qualtrics survey to assess which cue they thought was the hardest, easiest, most helpful, and least helpful. We also collected demographic information (e.g., age, gender, AR/VR experience, etc.).

## IV. RESULTS

Outlier criteria were based on whether accuracy was at or below chance performance for all cueing conditions. Of the 57 participants, one participant's data was removed from the analysis. First, we analyzed the data in R using 3 (cue type) by 2 (reliability level) X 2 (cue location) mixed ANOVAs to examine the effect of cue type (arrow, minimal, icon, and no cue), cue reliability level (100% and 83%) and location (up vs down) on performance. Cue type was submitted as the within-subjects factor and reliability level as the between-subjects factor. We log-transformed response time data because the assumptions of normality and homogeneity were violated.

### A. Overall Cueing Effectiveness

*Response Time*. The effects of cue type and reliability level on response time (measured in seconds) are shown in Figure 3. To ease the interpretation, we plotted raw response time rather than the log-transformed response time. The single value of mean response time in the control (no cue) condition is plotted as the red dashed horizontal line, the red shaded region represents 1 standard error of the mean.

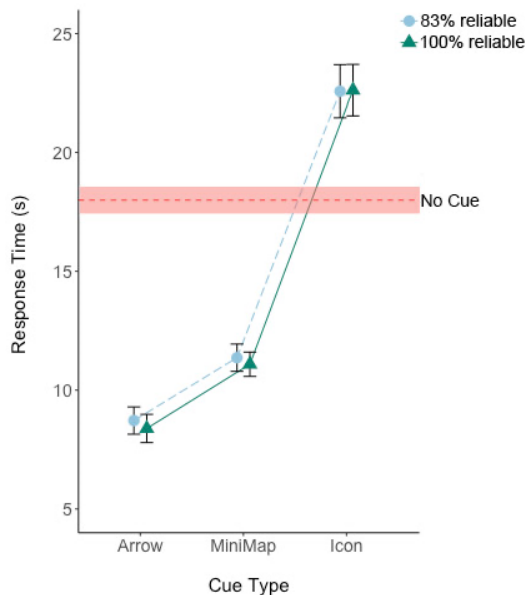

**Figure 3.** Mean response time plotted as a function of cue type and reliability. Light blue dashed and green solid represent 83% and 100% reliability, respectively. Error bars represent the standard error of the mean. The horizontal dashed red line represents the mean response time for the no cue (control) condition. The shaded region indicates the standard error of the mean.

To examine $H_1$, the overall benefits of cueing to response time, separate planned comparisons were made between the no cue and the three cueing conditions. These revealed significant benefits of both the arrow cue ($t(55)$ = -8.69, $p < .001$, $d$ = 1.45) and the minimap cue ($t(55)$ = -5.66, $p < .001$, $d$ = 0.87) with large effect sizes, but participants were actually slower than the uncued condition in finding the target with the icon cue ($t(55)$ = 3.73, $p$ = .0005, $d$ = 0.44). Thus, $H_1$ was supported in two of the three cases.

To examine $H_2$, specifically regarding the differences in cue effectiveness for response time, a 3 (cue type) by 2 (reliability), the mixed ANOVA was carried out on the log-transformed response time. The assumption of sphericity was violated, therefore Greenhousse-Geisser corrections for log-transformed response time data are reported here. This ANOVA revealed a significant, large effect of cue type on log-transformed response time, $F(1.52, 82.13)$ = 105.89, $p < .001$, $\eta_p^2$ = 0.66. As suggested by the similarity of the two curves in Figure 3, the reliability level did not impact log-transformed response time, $F(1, 54)$ = 0.05, $p$ = .82, $\eta_p^2 < 0.1$. The interaction was also not significant ($p$ = .42). Specific contrasts revealed, as shown in Figure 3, that all three cueing conditions differed from each other, with the arrow supporting fastest performance, and the icon supporting slowest performance. As the figure suggests, the minimap was also much more effective than the icon ($t(55)$ = 11.01, $p < .001$, $d$ = 1.40), although still significantly slower than the arrow cue ($t(55)$ = -6.30, $p < .001$, $d$ = 0.82). Thus, these results fully support the ordering of cueing effectiveness on search speed proposed in $H_2$.

*Accuracy*. Figure 4 presents the accuracy data in identical format to the response time data of Figure 3. In contrast to the response time data, these data indicate that all six cueing conditions, whether perfectly reliable or not, improved accuracy over the no cue (control) condition, supporting $H_3$.

The 3 (cue type) by 2 (reliability) mixed ANOVA conducted on the accuracy data revealed a significant effect of cue type, $F(2, 108)$ = 4.83, $p < .01$, $\eta_p^2$ = 0.08. In addition, there was a significant and large effect of reliability level on percent error, $F(1, 54)$ = 27.51, $p < .001$, $\eta_p^2$ = 0.34, discussed below. There was no significant interaction between cue type and reliability level, $F(2, 108)$ = 0.24, $p$ = .78, $\eta_p^2 < 0.01$.

The analysis revealed that, as with response time, the arrow cue was most effective in supporting accuracy. However, the ordering for the icon and minimap cue was reversed from that shown for response time, with the icon cue showing somewhat, although not significantly, greater accuracy than the minimap cue, ($t(55)$ = 1.46, $p$ = .15, $d$ = 0.20). Thus, $H_4$ is supported, replicating the general findings of [24].

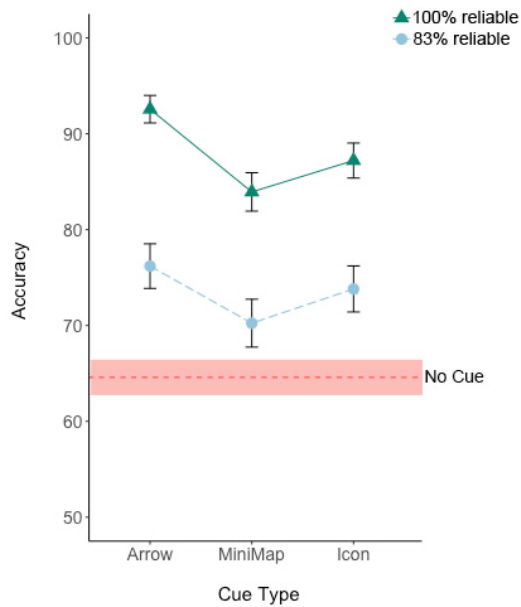

**Figure 4.** Mean accuracy plotted as a function of cue type and reliability. Green solid and light blue dashed represent 100% and 83% reliability, respectively. Error bars represent the standard error of the mean. The dashed red line represents the mean accuracy for the no cue (control) condition. The shaded region indicates the standard error of the mean.

### B. The Automation Bias: Automation Error Trials

The data addressing $H_5$, the effects of imperfection were described in the two analyses above. In the analysis of RT, imperfection of cueing had no effect. In the analysis of accuracy, consistent with $H_5$, it had a large effect, degrading accuracy across all cue types by approximately 13%. This degradation could result from either or both of two causes: (1) following the automation bias, participants blindly followed the cue, whether right or wrong, and hence, when it was wrong, suffered the corresponding increasing error rate; (2) participants, knowing that the cue could be wrong on any trial, relied less upon the automation, and more on their own imperfect performance level. Because when they were unaided, this performance level at 65% (see horizontal line of Figure 4) was less than with the cueing aids, such increased self-reliance would also pull the overall average performance downward.

The data to examine the automation bias are presented in the speed-accuracy tradeoff space in Figure 5. In this, we have plotted the data for all three conditions of cue effectiveness when the cues were accurate and when they were wrong. Our statistical analysis is performed only on the two most different conditions in cueing effectiveness when the cue was correct, specifically the arrow and the icon cue. These two represent very different information processing mechanisms (see equations (2) and (4)). Inspection of the figure also reveals that performance with the two location cues, the arrow and minimap, respond almost identically to the failure.

Each line in Figure 5 connects the three different cue types: green triangular arrows and blue circles toward the left, reflect a rapid RT and the square red icon toward the right reflecting their longer RT. Each data point represents the speed (RT) and accuracy of **correct cueing** (top) and **wrong cueing** (bottom).

The larger standard error bars for the wrong-cueing trials result from the fact that their N was 1/6$^{th}$ the size of the N for correct automation. For comparison purposes, the unfilled circle represents the data from the no cue (control) condition.

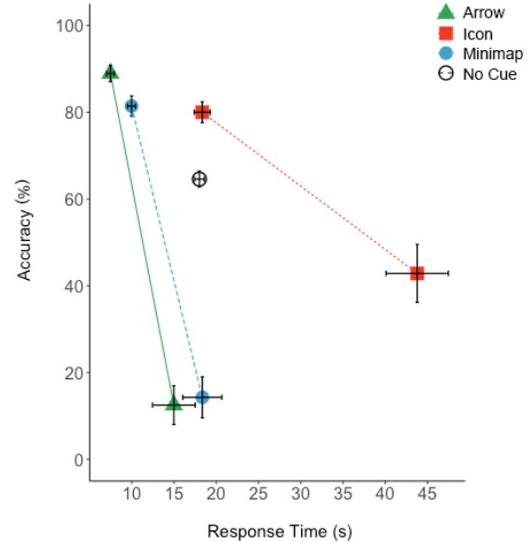

**Figure 5**. Performance in the speed-accuracy space. Mean response time (x-axis) by mean accuracy (y-axis) for the arrow (green triangle, solid line), minimap (blue circle, dotted line), and icon (red square, dashed line) cues when the cue was correct (top three data points) and incorrect (bottom three data points). The open circle represents the no cue (control) condition. Horizontal and vertical error bars represent the standard error of the mean for response time and accuracy, respectively.

Within this speed-accuracy space, high cue effectiveness is represented in the upper left corner: short RT and high accuracy. The data when the cue is correct, show the pattern described above in Figures 3 and 4: the arrow cue is unambiguously more effective on both speed and accuracy variables. However, when the cue is wrong (lower data points), the pattern essentially reverses. There is a delay in all cue types, suggesting that participants are somewhat or sometimes aware of the cueing error (i.e., not always blindly following the cue, but sometimes just more hesitant in responding, which represents an increase in the constant $a$ in equations 2-4). There is a large loss in accuracy for all three cue types suggesting an automation bias. Critically, however, this loss is much greater for the arrow cue, where accuracy drops from 89% to 13%, than it is for the icon cue, where accuracy drops from 80% to 43%. That is, participants correctly chose the blue box (target not present) on 57% of the trials. The significant difference in accuracy between the two cue types when the cue is wrong showed a large effect size, ($t(27)) = 9.50$, $p < .001$, $d = 2.59$), thereby supporting $H_6$ It should be noted here that the positioning of the minimap cue in the speed-accuracy space when the cue is wrong is nearly identical to that of the arrow cue.

Above, we described the two different kinds of human errors that could occur when the automation was wrong: the participant's decision to use their fallible judgment, and the blind following of the automation (the automation bias). The prevalence of the former can be estimated from the error rate in the control condition, which is approximately 30%. Hence, we can assume that of the 60% error rate in the icon condition half of these can be attributed to the *automation bias errors*, and for the two location cue conditions, the approximate 92% error rate can be predominately attributed to the *automation bias* errors. The reduced automation bias of the icon condition would appear to reflect the more thorough examination of the target object during the confirmation stage, and hence the longer RT delay as reflected by the constant $a' > a$.

### C. Location Effects

In addition to our primary interest in cue type and reliability, a secondary interest included the influence of cue location. Our prior research [24] indicated that performance was somewhat disrupted due to increased visual scanning between the object view and the cue view when the cue was positioned downward. In the current experiment, we were also interested in any possible interaction between cue location and the location of the target object in the wider visual field. The latter variable could provide us insight into the visual search strategies employed in unaided search and how they might be modified by the different cueing types. To conduct the analyses of both cue and target location effects, we separated target location by its vertical (shelf, table, floor; see Figure 1) and lateral positioning. Because the targets were continuously distributed in the lateral field, we coded lateral location into four categories: left versus right of initial fixation, and toward the center or the periphery. In this way we could discern one of two different search strategies often employed: left-to-right, and center-outward. We analyzed the three levels of vertical target positioning, separately from the four levels of lateral positioning, to ensure sufficient data in each. We also analyzed the uncued trials separately from the cued trials, only the latter analyses including the variable of cue location.

The effects of both cue and target location were, for the most part, muted and non-significant. In particular, cue location (center or downward) showed no significant effects nor interactions with either RT or accuracy.

The only prominent effects of vertical target location were for RT. Results showed faster RT for the top ($M_{shelf} = 13.43$ seconds) and middle ($M_{table} = 13.65$ seconds) locations than the bottom ($M_{floor} = 15.29$ seconds), $F(2, 110) = 9.45$, $p < .001$, $\eta_p^2 = 0.15$, suggesting some combination of a top to bottom search strategy, and a middle (table) outward (to shelf or floor) strategy. This effect was also significant, but smaller in magnitude and of the same pattern for uncued trials, $F(2, 110) = 3.62$, $p = .03$, $\eta_p^2 = 0.06$. There was no consistent trend for lateral position in either cued or uncued trials.

For accuracy (on the cued trials), there was an interaction between vertical location of targets and cue location, $F(2, 110) = 5.06$, $p = .008$, $\eta_p^2 = 0.08$. For the floor objects, greatest accuracy occurred when the cue was presented downward ($t(55) = 2.92$, $p = .005$, $d = 0.50$), a benefit for downward cueing not conferred in other vertical target locations. The effect of lateral target location (on cued trials only) showed greater accuracy when the target was in the center, $F(1, 55) = 8.42$ $p = .005$, $\eta_p^2 = 0.13$. There were no consistent effects of lateral target position on accuracy for uncued trials. The absence of any main effect of cue location fails to confirm H7.

## V. DISCUSSION

In this experiment, we have joined two HCI issues in the study of AR-HMD target cueing: the capability of HMD cueing to exploit properties of augmented reality to offer world-referenced cues and the effects of imperfect reliability of automation-generated cues.

### A. Overall Cueing Effectiveness

In a task requiring the integration of near domain (displayed cue) information and far domain (search field) information, we found AR cueing (i.e., the arrow) to provide a pronounced 10 second benefit relative to uncued targets (H1). We also found a benefit for cues that were not presented in world-referenced augmented reality coordinates compared to uncued targets. Specifically, the minimap showed a 4 second benefit and the icon cue showed a 15 % accuracy benefit relative to no cue, confirming H2.

The time benefit of AR arrow cueing over the minimap was small (2 seconds) but significant. Both techniques provided precise **location** cueing, thus eliminating the $nT/2$ from predictive Equations 2 and 3. However, the minimap cue imposed some added time cost due to the frame-of-reference transformation, when transforming a top-down 2D map to the forward 3D view of the search scene [31], [49]. This penalized the CEV by approximately 2 seconds (Equation 3). While no spatial transformation was required for the icon cue, its time penalty relative to the arrow cue was quite large, at approximately 15 seconds: a 200% increase.

In the context of the serial self-terminating model of visual search discussed in the introduction [20] assume that, with the icon cue, each object is examined and compared with the icon cue in order to establish that it is not the target. This search strategy continues until a match is found, a comparison process of time T, that we estimate from the current data to be about 0.40 seconds. In the context of the equations presented in the introduction, this estimate assumes that, for the precise location cues, only one comparison is required (N = 1), whereas for the icon cue, on average, $48/2 = 24$ comparisons are required. (We note that in [24], the estimate of T was quite similar – approximately 0.33 seconds – to that observed here 40 seconds). The much longer search time of the icon cue could be due to some combination of a longer confirmation time $(a' > a)$ and the need to inspect many more images: $((n - x)T)/2$. The current data do not allow us to determine the precise value of CEV for the icon cue because we have no way to estimate "$x$", the number of items that do not require inspection because of guided search [21].

While this model accounts for the differences in search time between the cueing conditions, and supports the advantage of location over identity cueing, the accuracy differences result from a different mechanism. In particular, the accuracy of the icon was now elevated to be highly similar to that of the minimap, and non-significantly greater. The reason for this equivalence is assumed to lie in the target confirmation process (a process reflected in the RT function by the intercept $a$ or $a'$). For the icon, this confirmation is a direct **visual** comparison between the viewed target object and the icon. However, for the minimap, it is a comparison between the visual image of the target, and the **remembered** image of the cue, which was viewed at the beginning of the trial. For the minimap, this time delay, averaging approximately 12 seconds (see Figure 3), is of sufficient magnitude such that it degrades the quality of the image in spatial working memory [56] and produces some loss to the accuracy of comparison. This loss for the minimap is greater than it is for the arrow cue simply because the minimap time-delay is about 3 seconds longer than that of the arrow (see Figure 3). That time difference is great enough to account for an accuracy decrement of the minimap relative to the arrow in the quality of the remembered target image. With a lower quality image, accuracy will be compromised. We note that in all cases, the accuracy is far below perfect. That is because the Mega Block targets were intentionally designed to be highly similar to many of the non-targets in their 3D shape and color combination (see Figure 2 right panel). This might represent, for example a vehicle with a certain differentiating feature (e.g., 1 foot longer) compared to many other non-target vehicles sharing many features to the target (e.g., shape, color, number of doors, etc.).

Despite the pronounced advantage of location (arrow and minimap) over appearance (icon) cueing, it is possible to envision circumstances in which the advantage is reduced or even reversed. For example, in some circumstances initial location uncertainty may be quite small, as in searching a line-up of only five (N = 5) suspects for a crime to identify the remembered target previously viewed at the crime scene; but there may be subtle differences in facial appearance of the true target and the non-targets, greatly impinging on accuracy, when there is no image for direct visual comparison.

*B. Imperfection of Cueing*

The second intersecting issue was the 17% imperfection of AI, which degraded overall human-automation team accuracy and CEV by 15% supporting H5. This effect of automation reliability has been well documented in much of the HAI research in general [57] as well as that specific to target search [15], [36], [37], [10]. These findings indicate that imperfection of a cue will almost always pull human-automation team performance downward relative to perfect cueing, as long as accuracy performance (reliability) of the human alone is less than that of the imperfect cue alone, as was the case in the present experiment (Figure 4). But at the same time, imperfect reliability in automation will typically aid human performance accuracy above an unaided manual baseline, as is the case here ([36], [37, [47], see Figure 4). In this context it is also important to note that even with the best cue of the perfect cueing condition the mean accuracy of 92% is still less than that had participants only followed the cue which was correct 100% of the time. Such a finding is not unexpected, given that other studies have observed that performance with highly reliable automation remains below that of automation itself ([36], [37], see [2] for a summary).

Our more detailed examination of the data indicated that the consequence of cueing imperfection was not simply a blind adherence to the automation bias [41], [42]. If this had been the case, then the speed of responding would have been unaffected by whether the cue was or was not correct. But, as shown in Figure 5, speed was somewhat affected. The response was significantly slower on those occasions when the cue was wrong, even for the most effective cue (i.e., the arrow), which might have been predicted to show the greatest automation bias. We suspect that such hesitation may have been the result of increased overall caution with the participant knowing that cueing errors could occur, or the increase in confirmation time on those trials where doubt might have occurred in the participant's mind. This would produce an increase in the search time (RT) constant $a$ in equations (2) and (3), or $a'$ in equation (4)

Nevertheless, strong evidence was provided for the presence of the automation bias in all cueing conditions. It is here that the two HAI issues, that of AR and automation imperfection, intersect. When location cueing is most intuitive (i.e., with the AR arrow) in world-referenced coordinates, it is most likely to be followed without using the raw data of the target object in the search scene, supporting H6. Mifsud and colleagues [30] also found this automation bias with a different form of AR target cueing, and [14] found it to be the case when comparing AR-HMD cueing with non-AR cueing. As described in the Introduction above, the more realistic the cue, and perhaps the more powerful the response it elicits the more likely it is to be followed when it is wrong: for example, in this case, pointing an arrow at a potential target that the user must then designate for the response. This can be seen as a specific manifestation of the more general "Lumberjack" tradeoff principle of human-automation interaction: the more effective automation is when it is correct, the greater the consequences to the human-automation team's performance when it is wrong [39]. (That is, the higher the tree, the harder it falls). This leads us to consider two aspects of human-automation interaction. On the one hand, to consider mitigating solutions to over-dependence on automation and AI when its information is rendered in AR. On the other hand, to consider the frequency and kind of errors committed by AI when it is based on actual machine vision operating in real-time under time-sensitive conditions of potentially hostile behavior detection or dynamic hazard avoidance [45].

Our findings regarding location effects were muted. Differences in the location of the cue, center or downward on the HMD, had little effect on performance. Thus, the information access penalty of a small downward scan to see any of the cues was quite small, similar to findings in another context [53]. It may be the case that the downward position of the cue location (at 12.6 degrees) remained sufficiently within the eye-head field range to mitigate any cost of information access effort that we might otherwise see if the cues were out of the field-of-view of the HMD entirely (e.g., glanceable AR; [58]). Correspondingly, there was not the predicted interaction of cue location with target position that might have suggested a clutter effect. This clutter effect refers to the greatest performance decrement in two circumstances: (1) when the cue was in the center and the target on the table, and (2) when the cue was positioned downward, and the target was on the floor. This clutter effect was also muted, possibly because the minimap was a smaller size than the prior study [24], creating less overlay clutter. Alternatively, the performance penalties for increased clutter due to overlaying target and cue were offset by the benefits of reduced information access between target and cue location.

In this regard, the significant interaction observed between cue and target vertical location suggested that the lower targets were best served by the bottom cue location: close spatial proximity between cue and target fosters best performance. This finding is a prediction of the proximity compatibility principle [59], [53], which states that when two sources of information require integration, those sources should be placed closer together in space.

## VI. Limitations

One limitation of the current research, in generalizing to cued search in the real world, is that the latter includes many cases in which the objects need not be compared with an image to be detected. For example, consider searching for a weapon in an X-ray image at airport security. The searcher has a well-formed image in long term memory of what a "typical weapon" looks like. Yet the aforementioned example of search for a particular vehicle that may have a similar shape and color of other vehicles but differs in its length, or instances where specific targets may be less familiar and are not stored in long-term memory, demonstrate cases more typical of that studied here where objects cannot be easily discriminated from foils.

Another limitation is that the arrow, like any AR cue, could be subject to parallax errors. The magnitude of this effect may be quite dependent on the proximity between the target and adjacent items. If search items are close together, parallax errors could easily result in an arrow cue pointing to an adjacent item rather than the true target. In the current research, we assured sufficient spacing between search items that this did not occur; but it certainly remains a concern in other applications.

Lastly, we did not use expert participants who engage in visual search tasks, such as those in a military combat role for which this paradigm might be highly relevant. We simply did not have a sufficient number of expert participants available to gain adequate statistical power. However, we note that in other AR-HMD studies, our subject matter experts have stated that a major concern of the JTAC is that of confusing a relevant from a non-relevant object on the ground when searching for the former. Therefore, we are of the firm belief that the current research addresses a highly relevant applied issue.

## VII. Conclusions and Future Work

Overall, we found that cues depicting location information, the arrow and minimap cues, led to faster and, in the case of the arrow cue, more accurate visual searches compared to the icon cue, which only depicted information about the appearance of a target. However, when imperfect cues were erroneous, the most effective cues for supporting search performance led to worse outcomes. This is indicative of an automation bias where people blindly follow the cue without sufficient, independent confirmation for themselves that the identified object is the correct target. These findings highlight an important prospective use case for AR support during visual search tasks, and also shed light on the costs associated with imperfectly reliable cueing aids. A tendency towards greater automation bias may be a critical drawback in some high-stakes contexts. Such findings strongly imply that designers will need to take a system view of error management in supporting visual search rather than just optimizing performance under an assumption of perfectly reliable support.

In the future, several experiments are warranted to explore the interaction between the two HMD automation issues. For example, different levels of automation reliability should be explored to assess if the augmented reality cost for the more realistic world-reference cue diminishes as cueing reliability also decreases. Also, the current findings when automation errors are scripted by the experiment should be extended to circumstances in which actual machine vision determines target cueing in real-time. Correspondingly, effort should be taken to systematically manipulate target versus non-target similarity, to establish if advantages to accuracy of the icon cue begin to emerge, particularly as the size of the search field, *n,* is reduced. Finally, more research should be conducted when the screen-referenced cue is further displaced from the search field, thereby increasing information access effort.